\documentclass[conference]{IEEEtran}
\usepackage{algorithm, algorithmic}

\usepackage{tabularx}
\usepackage{array}
\usepackage{ifpdf}
\ifCLASSINFOpdf
\else
\fi
\usepackage{url}
\usepackage{amsmath,graphicx}
\usepackage{placeins}
\usepackage{amsmath,epsfig}
\usepackage{epstopdf}
\usepackage{amssymb}

\usepackage{dblfloatfix}
\usepackage{amssymb}
\usepackage{caption}
\usepackage{comment}
\usepackage{subfigure}
\usepackage{pstricks}
\usepackage{subfloat}
\begin{document}
%\title{Low Complexity Algorithm Based on Proximal Jacobian ADMM For Uplink Massive MIMO Signal Detection in URLLC}
\title{A Proximal Jacobian ADMM Approach for Fast  Massive MIMO Signal Detection in Low-Latency Communications}

\author{\IEEEauthorblockN{}
\IEEEauthorblockA{*Anis Elgabli, Ali Elghariani, Vaneet Aggarwal, *Mehdi Bennis, and Mark R. Bell\\
Purdue University, West Lafayette IN, USA, * University of Oulu, Finland \\
}}
\maketitle
%\input{sec01_intro}
%\vspace{-.4in}
\begin{abstract}
One of the 5G promises is to provide Ultra Reliable Low Latency Communications (URLLC) which targets an end to end communication latency that is $\leq$ 1ms . The very low latency requirement of URLLC entails a lot of work in all networking layers. In this paper, we focus on the physical layer, and in particular, we propose a novel formulation of the massive MIMO uplink detection problem. We introduce an objective function that is a sum of strictly convex and separable functions based on decomposing the received vector into multiple vectors. Each vector represents the contribution of one of the transmitted symbols in the received vector. Proximal Jacobian Alternating Direction Method of Multipliers (PJADMM)  is used to solve the new formulated problem in an iterative manner where at every iteration all variables are updated in parallel and in a closed form expression. The proposed algorithm provides a lower complexity and much faster processing time compared to the conventional MMSE detection technique and other iterative-based techniques, especially when the number of single antenna users is close to the number of base station (BS) antennas. This improvement is obtained without any matrix inversion. Simulation results demonstrate the efficacy of the proposed algorithm in reducing detection processing time in the multi-user uplink massive MIMO setting.
\end{abstract}


\section{Introduction}
\label{sec:intro}

Massive  multiple-input  multiple-output  (MIMO)  is  one of  the  most  promising  techniques  in 5G networks due to its potential for significant rate enhancement and energy efficiency \cite{boccardi2013five}, \cite{van2017massive}. An important requirement  for the emerging URLLC 5G services is to provide low latency and fast processing time for these services \cite{bennis2018ultra}, especially in the multi-user massive MIMO scenario, in which the number of users is greatly increasing \cite{li20185g}. The challenge of providing the low latency and fast processing is approached in the literature through several aspects and at different network layers \cite{{anand2018resource}, {shapin2018physical}}. For example, in \cite{bacstuug2014living} edge caching and high computation edge nodes are discussed as ways to reduce latency. In physical layer, and in particular in the resource allocation problem, Non Orthogonal Multiple Access (NOMA) in combination with grant-free can be used in URLLC \cite{tian2017uplink}. On the other hand, for a waveform design problem \cite{ashraf2017radio} presents a choice of suitable modulation and coding schemes, and discuss the impact of different waveform candidates. In this paper, we focus on the signal detection component of the receiver, in which we propose a fast signal detection algorithm based on parallel implementation. 

Reducing the computational complexity of the detection algorithm is very important since the less the number of computations the algorithm needs, the faster processing time it requires. Thus, from this perspective, to reduce the computational complexity, linear MMSE detection scheme is widely considered, 
%which rely on a truncated Neumann series expansion, have been proposed in [6]?[9]. This approach requires (often significantly) lower computational complexity than that of an exact inversion while delivering near-optimal results for massive MIMO systems
but this scheme involves matrix inversion which is computationally costly.  A truncated Neumann series expansion can be used to approximate the matrix inversion \cite{wu2014large}. With sufficient large truncation order, the approximation can be very close to real matrix inversion only in cases when the ratio between the number of BS antennas and the number of single antenna users is large (e.g.  $\geq $16  ). However, \cite{wu2014large} indicates that when the ratio becomes smaller, a large truncation order is required which makes the computational complexity more than the exact matrix inversion operations. Similar other ideas are presented in \cite{yin2014conjugate} focusing on how to reduce the complexity of the matrix inversion, but they suffer from the deficiency  of deteriorated performance and increased computations as the number of single antenna users increases \cite{yin2014conjugate}.

Another approach is based on iterative methods to reduce computational complexity. The iterations are implemented to transform the matrix inversion problem of MMSE matrix into solving linear equations \cite{jiang2017low}, \cite{qin2016near}. The iterative methods can also be used in a context other than finding the inversion of the MMSE matrix, such as \cite{elgabli2018low}, \cite{elghariani2016low}, where the detection problem is formulated as a convex optimization problem that can be solved in an iterative manner using alternating minimization techniques or quadratic programming techniques.

The above two approaches experience long waiting delay, especially in the case of a large number of users. For instance, the information symbol belongs to the last user in the received signal vector needs to wait until all previous symbols pertaining to all other users are detected \cite{jiang2017low}. This successive detection manner makes the detection scheme time inefficient. Besides, this is not efficient for hardware implementation \cite{jiang2018low}.

Due to the increased interest for dealing with big data and large scale problems such as the problem at hand, parallel and distributed computational methods are highly desirable for faster processing time. Alternating Direction Method of Multipliers (ADMM), as a versatile algorithmic tool, has proven to be very effective at solving many large-scale problems and 
\\
well suited for distributed computing \cite{deng2017parallel}. In this paper, we, first, we use our propose a novel formulation of the uplink massive MIMO detection problem in \cite{elgabli2019altmin} in such a way that it fits the ADMM formulation. In particular, we introduce an objective function that is a sum of strictly convex and separable functions based on decomposing the received vector into multiple vectors. Each vector represents the contribution of one of the transmitted symbols in the received vector. Second, we use the Proximal Jacobian Alternating Direction Method of Multipliers (PJADMM) \cite{deng2017parallel} to solve the new formulated problem in an iterative manner such that in every iteration all variables are updated in parallel and in a closed form expression, without requiring any matrix inversion. Therefore, at every time instance, information symbols of all users are updated in parallel at the receiver. This in turn provides  much faster processing time compared to the conventional MMSE detection techniques, and also compared to the other iterative-based techniques, especially when the number of users (each with a single antenna) is close to the number of base station (BS) antennas. Note, unlike the existing techniques such as in \cite{jiang2018low}, our proposed algorithm does not try to parallelize existing MMSE detection problem, instead it develops the parallel implementation of the detection problem based on a unique formulation of the maximum likelihood optimization problem combined with the application of PJADMM technique. In the end, our objective is to reach the exact MMSE performance but with a faster processing time. Simulation results demonstrate the efficacy of the proposed algorithm in the multi-user uplink massive MIMO setting.

\indent The remainder of this paper is organized as follows. In Section II, the system model and problem formulation are described. Section III contains the proposed algorithm, and section IV presents simulation results, and finally the paper is concluded with a short summary.

\section{System Model and Problem Formulation}
\subsection{System Model}
Consider the uplink data detection in a multi-user massive MIMO system with $N_r$ BS antennas and $N_t$ users each with a single antenna. The vector $\tilde{\textbf{x} }= (x_1, x_2, \dots, x_{N_t})^T \in  \mathbb{C}^{N_t  \times1}$ represents the complex transmitted signal, where $x_k$ is the transmitted symbol for user $k$ with $E{|x_i|^ 2} = 1 , \forall i$. Each user transmits symbols over a flat fading channels and the signals are demodulated and sampled at the receiver. The vector $\tilde{\textbf{y}} = (y_1, y_2, \dots , y_{N_r} )^T \in \mathbb{C}^{{N_r} \times1}$ represents the complex received signal. The channel matrix $\tilde{\textbf{H}}  \in \mathbb{C}^{{N_r} \times N_t} $ can be represented as $(h_1, h_2, \dots, h_{N_t})$, where $h_i =(h_{1,i}, h_{2,i},..., h_{N_r,i})^T  \in \mathbb{C}^{{N_r} \times 1}$, and $h_{m,n}$ is the complex flat fading channel gain from transmit antenna $n$ to the receive antenna $m$, with $h_{m,n}  \sim \mathcal{CN}(0,1)$. The system can be modeled as: %$\tilde{\textbf{{y}}}=\tilde{\textbf{{H}}} \tilde{\textbf{{x}}}+\tilde{\textbf{{v}}}$,
\begin{equation}
\tilde{\textbf{{y}}}=\tilde{\textbf{{H}}} \tilde{\textbf{{x}}}+\tilde{\textbf{{v}}}
\label{eq: 1}
\end{equation}
where $\tilde{\textbf{v}} = (v_1, v_2, иии , v_{N_r} )^T \in \mathbb{C}^{{N_r} \times N_t} $ is the complex additive white Gaussian noise (AWGN) vector whose elements are mutually independent with zero mean and variance $\sigma^2_v$. %We assume that the transmitted signals are i.i.d Gaussian distribution and satisfy $E{|x_i|^ 2} = 1 , \forall i$, i.e. The symbols are normalized to unit energy.
\if0
\begin{figure}[htp]
%%%\begin{minipage}[b]{1.0\linewidth}
 \centering
\includegraphics[width=6cm, height=4cm]{systemmodel_uplink.pdf}
\caption{\small  System Model for Massive MIMO uplink}
\label{fig:1}
\end{figure}
\fi
The corresponding real-valued system model is $\textbf{{y}}=\textbf{{H}}\textbf{{x}}+\textbf{{v}}$ \cite{elgabli2017two}. The equivalent ML detection problem of the real model can be written in the form $\widehat{{\textbf{x}}}=  \underset{{{\textbf{x}}\in\chi^{2N_t}}}{\text{argmin}} \parallel{{\textbf{y}}}-{{\textbf{H}}}{{\textbf{x}}}\parallel_2^{2}$,
where $ \chi=\frac{1}{\Gamma}\{-\sqrt{\textit{M}}+1,..,-1,1,...,\sqrt{\textit{M}}-1\}$, $\textit{M} \normalsize$ is the QAM constellation size, and $\frac{1}{\Gamma}$ the normalization factor.\\

\if0
\small
\begin{equation}
\widehat{{\textbf{x}}}=  \underset{{{\textbf{x}}\in\chi^{2N_t}}}{\text{argmin}} \parallel{{\textbf{y}}}-{{\textbf{H}}}{{\textbf{x}}}\parallel_2^{2}
\label{eq: 6}
\end{equation}
\normalsize
\fi

\subsection{Problem Formulation}
First, we decompose the received vector $\textbf{y} $ into a linear combination of vectors so that $\textbf{y} = \sum_{i=1}^{2N_{t}}{\textbf{y}_i}$, where $\textbf{y}_i$ represents the contribution of the $i$-th transmitted symbol in the received vector. The element wise representation of the decomposed received vector is: 
\small
\\
\\
$
\begin{bmatrix}
y^{(1)}\\
y^{(2)}\\
.\\
.\\
.\\
y^{(2N_r)}\\
\end{bmatrix}
=
\begin{bmatrix}
y_1^{(1)}\\
y_1^{(2)}\\
.\\
.\\
.\\
y_1^{(2N_r)}\\
\end{bmatrix}
+\cdots+
\begin{bmatrix}
y_i^{(1)}\\
y_i^{(2)}\\
.\\
.\\
.\\
y_i^{(2N_r)}\\
\end{bmatrix}+\cdots+
\begin{bmatrix}
y_{2N_{t}}^{(1)}\\
y_{2N_{t}}^{(2)}\\
.\\
.\\
.\\
y_{2N_{t}}^{(2N_r)}\\
\end{bmatrix}
$
\\
\\
\\
\normalsize
The  $k$-th element of $\textbf{y}$ can be represented as ${y}^{(k)} = \sum_{i=1}^{2N_{t}}{{y}_i^{(k)}}$, $k=1,\dots, 2N_r$. Let $\textbf{h}_i$ the $i^{th}$ column of the real channel matrix $\textbf{H}$.  Now, we relax the non-convexity constraint on the feasible set $\chi$, and approximate the ML problem based on the above decomposition as follows: 
\begin{equation}
\underset{{x_i},\textbf{y}_i}{\text{min}}\sum_{i=1}^{2N_{t}} \parallel{{\textbf{y}_i}}-{{\textbf{h}_i}}{{{x}_i}}\parallel_2^{2}
\label{eq: 7}
\end{equation}
\indent \,\,\,\,\,\, subject to
\begin{equation}
\sum_{i=1}^{2N_{t}}{{y}_i}^{(k)}={y}^{(k)}, \forall k=1,\cdots,2N_r
\label{eq: 7c1}
\end{equation}
\begin{equation}
-l \leq {x}_i \leq l, \forall i=1,\cdots,2N_t
\label{eq: 7c2}
\end{equation}
where $l= \frac{1}{\Gamma} (\sqrt{\textit{M}}-1)$. The objective function in (\ref{eq: 7}) is a sum of separable terms, each of which is a function of only one symbol and its contribution in the received vector. In the next section, we use AltMin to solve the proposed formulation.

\subsection{K.K.T Conditions}
 
 The problem in~\eqref{eq: 7}-\eqref{eq: 7c2} is a convex optimization problem with linear constraints, so the K.K.T. conditions are sufficient and necessary for the optimal solution.%, we use the K.K.T. conditions
%to characterize the optimality of the problem in~\eqref{eq: 7}-\eqref{eq: 7c2}.
 The Lagrangian function for~\eqref{eq: 7}-\eqref{eq: 7c2} is defined as follows:
 
 \begin{align}
&\mathcal{L}=\sum_{i=1}^{2N_{t}} \parallel{{\textbf{y}_i}}-{{\textbf{h}_i}}{{x_i}}\parallel_2^{2}+\sum_{k=1}^{2N_r}\lambda^{k}(y^{(k)}-\sum_{i=1}^{2N_{t}}{y_i}^{(k)})\nonumber\\&+\sum_{i=1}^{2N_t}\big(\mu_1^{(i)}(l-x_i)+\mu_2^{(i)}(l+x_i)\big)
\end{align}

Then, the following K.K.T. conditions, which are sufficient and necessary for the optimal solution to the convex optimization problem in~\eqref{eq: 7}-\eqref{eq: 7c2}, are obtained from the Lagrangian function:

\begin{equation}
y_i^{(k)}=h_i^{(k)}x_i+\lambda^{(k)}/2, \forall i,k
\label{eq: 10}
\end{equation}

\begin{equation}
\lambda^{(k)}=\frac{1}{N_t}(y^{(k)}-\sum_{i=1}^{2N_{t}}h_i^{(k)}x_i), \forall i,k
\label{eq: 11}
\end{equation}

\begin{equation}
2x_i\sum_{k=1}^{2N_t}{h_i^{(k)}}^2-2\sum_{k=1}^{2N_t}y_i^{(k)}h_i^{(k)}-\mu_1^{(i)}+\mu_2^{(i)}=0, \forall i,k
\label{eq: 131}
\end{equation}

\begin{equation}
\mu_1^{(i)}(l-x_i)=0, \forall i
\label{eq: 141}
\end{equation}

\begin{equation}
\mu_2^{(i)}(l+x_i)=0, \forall i
\label{eq: 151}
\end{equation}
 
 %\textcolor{green}{However, solving the K.K.T. conditions to find the optimal solution in closed form expression requires high complexity computations. Therefore, in next section, we propose our iterative algorithm based on PJADMM where all variables are updated in closed form expression per iteration. Moreover,  transmitted symbols of different users and their corresponding contributions to the received vector are updated in parallel. We do not discuss the optimality of the proposed algorithm for two reasons i) First, the optimality of PJADMM was proven when the objective function is sum of strictly convex separable functions which is the case of our objective function \cite{deng2017parallel}, ii) second, because of the lack of space.}
 
In order to solve K.K.T. conditions to find the optimal solution in a closed form expression, high complexity computations are required. Therefore, in the next section, we develop our iterative algorithm based on the PJADMM in which all variables are updated in a closed form expression per iteration. That is, transmitted symbols of different users and their corresponding contributions to the received vector are updated in parallel.
 
\section{Proposed Algorithm based on Proximal Jacobian ADMM}

In Proximal Jacobian ADMM all blocks of the variables are updated in parallel. Proximal Jacobian ADMM was shown to enjoy a global convergence with convergence rate of $o(1/t)$ \cite{deng2017parallel}. One difference of Proximal Jacobian compared to Gauss Seidel ADMM is that the term $\frac{\tau}{2} \sum_{i=1}^{2N_t}(x_i-{x_i}^{(t-1)})^2+\frac{\tau}{2}\parallel \textbf{y}_i-{\textbf{y}_i}^{(t-1)}\parallel_2^2$ is added in the augmented Lagrangian. Hence, at every iteration, the primal variables, $\textbf{y}_i$, $x_i$ $\forall i$, are updated as follows:

\begin{equation}
{x}_i,\textbf{y}_i=\underset{{{{x}_i,\textbf{y}_i}}}{\text{argmin}}\mathcal{\boldsymbol{\mathcal{L}}}_{\rho, \tau}\big({x}_i,\textbf{y}_i, ({x}_j^{(t-1)},\textbf{y}_j^{(t-1)})\forall j \neq i, {\lambda}^{(t-1)}\big)
\label{augmentedLag4}
\end{equation}
In order to update $\textbf{x}$, the following optimization problem is solved.
\begin{align}
&{x}_i,\textbf{y}_i=\underset{{{{x}_i,\textbf{y}_i}}}{\text{argmin}} \parallel{{\textbf{y}_i}}-{{\textbf{h}_i}}{{x_i}}\parallel_2^{2}-\sum_{k=1}^{2N_r}{\lambda^{k}}^{(t-1)}{y_i}^{(k)}\nonumber\\&+\frac{\rho}{2}\sum_{k=1}^{2N_r} ({y}^{(k)}-y_i^{(k)} - \sum_{j=1, j \neq i}^{2N_{t}}{y_j^{(k)}}^{(t-1)})^2
\nonumber\\&+\frac{\tau}{2}({x}_i-{{x}_i}^{(t-1)})^2+\frac{\tau}{2}\sum_{k=1}^{2N_r}(y_i^{(k)}-{y_i^{(k)}}^{(t-1)})^2
\label{augmentedLag5}
\end{align}
subject to
\begin{equation}
-l \leq {x}_i \leq l
\label{augmentedLag3C1}
\end{equation}

To solve this problem, we first write the corresponding Lagrangian function of \eqref{augmentedLag5}-\eqref{augmentedLag3C1} which is:
\begin{equation}
\\
\text{eq}\eqref{augmentedLag5} + \mu_1^{(i)}(l-x_i) + \mu_2^{(i)}(l+x_i)
\label{lagrang_final}
\end{equation}
\\
\\
Now, we take the derivative of \eqref{lagrang_final} with respect to variables ${x}_i,\textbf{y}_i$ and set it to zero, for different combinations of  $\mu_1^{(i)}$ and  $\mu_2^{(i)}$, we get:

\begin{itemize}
\item  $\mu_1^{(i)}=0$, and $\mu_2^{(i)}=0$. Therefore, 
\begin{align}
&x_i=\frac{\frac{1}{2+\rho+\tau}\sum_{k=1}^{2N_r}h_i^{(k)}{\lambda^{(k)}}^{(t-1)}}{(1-\frac{2}{2+\rho+\tau}){\sum_{k=1}^{2N_r}h_i^{(k)}}^2-\tau}\nonumber\\&+\frac{\frac{\rho}{2+\rho+\tau}{\sum_{k=1}^{2N_r}h_i^{(k)}}(y^{(k)}-\sum_{j=1, j \neq i}^{2N_{t}}{y_j^{(k)}}^{(t-1)})}{(1-\frac{2}{2+\rho+\tau}){\sum_{k=1}^{2N_r}h_i^{(k)}}^2-\tau}\nonumber\\&+\frac{\frac{\tau}{2+\rho+\tau}{\sum_{k=1}^{2N_r}h_i^{(k)}{y_i^{(k)}}^{(t-1)}-\tau x_i^{(t-1)}}}{(1-\frac{2}{2+\rho+\tau}){\sum_{k=1}^{2N_r}h_i^{(k)}}^2-\tau}
\label{eq:x1Update}
\end{align}

\item  $\mu_1^{(i)}=0$, and $\mu_2^{(i)} \neq 0$. Therefore, 
\begin{equation}
{x}_i=-l
\label{eq:x1Update2}
\end{equation}  

\item  $\mu_1^{(i)}\neq 0$, and $\mu_2^{(i)} = 0$. Therefore, 
\begin{equation}
{x}_i=l
\label{eq:x1Update3}
\end{equation}  
\end{itemize}
There is still one more case in which  $\mu_1^{(i)}$ and $\mu_2^{(i)} $ both are not equal to zero, but it is void because $x_i$ cannot be equal to 1 and -1 at the same time.  Next, the update of $\textbf{y}_i$ is as follows:

\begin{align}
&{y}_i^{(k)}=\frac{1}{2+\rho+\tau}\Big(2h_i^{(k)}x_i^{(t)}+{\lambda^{(k)}}^{(t-1)}\nonumber\\&+\rho(y^{(k)}-\sum_{j=1}^{2N_t}{{y}_{j\neq i}^{(k)}}^{(t-1)})+\tau {y_i^{(k)}}^{(t-1)}\Big), \forall k
\label{eq:y1Update}
\end{align}  
The ideal case is to choose ${x}_i$ based on the above combinations of $\mu_1^{(i)}$ and  $\mu_2^{(i)}$ such that it minimizes the objective function \eqref{augmentedLag5}. However, since we perform hard quantization for detecting symbols, Eq\eqref{eq:x1Update} is enough. Finally, each dual variable ${\lambda^{(k)}}\forall k$ is updated as follows:

\begin{equation}
{\lambda^{(k)}}^{(t)}={\lambda^{(k)}}^{(t-1)}+{\rho}({y^{(k)}}^{(t)}-\sum_{i=1}^{2N_t}{y_i^{(k)}}^{(t)}), \forall k
\label{lambdaUpdate}
\end{equation}

The PJADMM algorithm is outlined in Algorithm 1. It is clear from ~\eqref{eq:x1Update}-\eqref{lambdaUpdate} that at every iteration, $t$, the elment $x_i^{(t)}$ depends on $\textbf{y}_i^{(t)}$, $x_i^{(t-1)}$, $x_j^{(t-1)}, \forall j \neq i$, $\textbf{y}_j^{(t-1)}, \forall j \neq i$ and ${\lambda^{(k)}}^{(t-1)} \forall k$. Therefore, it depends on the symbols' values of the previous iteration and their contribution to the received vector, and not on the current iteration values. The same is also true for the updates of $\textbf{y}_i$, and ${\lambda^{(k)}}$. This facilitates the implementation of the parallel processing at the receiver per every iteration, as shown in Algorithm~\ref{alg1}

\begin{figure}

		%\vspace{-.6in}
		\begin{minipage}{\linewidth}
			\begin{algorithm}[H]
					{\tiny 
				\small
				\begin{algorithmic}[1]
				 \STATE {\bf Input}: $\textbf{H}, \textbf{y}, \rho, \tau$
				  \STATE {\bf Output}: ${x}_i,\textbf{y}_i, \forall i$ 
				    \STATE {\bf Initilization}:  ${x}_i =0,\textbf{y}_i=\textbf{0}, \forall i, \lambda^{k} =0, \forall k$ 
    
   %\STATE $t=0$, $\textbf{y}_i=0\forall i$ 
%\STATE Solve~\eqref{augmentedLag5}-\eqref{augmentedLag3C1} to update $\textbf{x}_i\forall i$
%\STATE Solve~\eqref{eq:y1Update} to update $\textbf{y}_i\forall i$
%\STATE update $\lambda^{(k)}$ according to \eqref{lambdaUpdate}, $\forall k$

\STATE $\delta =$ convergence tolerance
\STATE $T=$ Maximum number of iterations

    \STATE {\bf Proximal Jacobian  ADMM:}
  %  \STATE { \bf  Parallel Loop}
    \REPEAT%\STATE {\bf REPEAT
 \STATE $t \leftarrow t+1$
 \STATE { \bf  Parallel Processing }
 \STATE use ~\eqref{eq:x1Update} to update $\textbf{x}_i\forall i$
\STATE use ~\eqref{eq:y1Update} to update $\textbf{y}_i\forall i$
\STATE use  ~\eqref{lambdaUpdate} to update $\lambda^{(k)}$, $\forall k$

\STATE $V^{(t)}=\sum_{i=1}^{2N_{t}} \parallel{{\textbf{y}_i}}-{{\textbf{h}_i}}{{\textbf{x}_i}}\parallel_2^{2}$
\UNTIL  $|V^{(t)}-V^{(t-1)}| < \delta$ {\bf OR}  $t > T$%\STATE {\bf UNTIL } $|V^{(t)}-V^{(t-1)}| < \delta$ {\bf OR}  $t > T$

   				\end{algorithmic}
				\caption{Proximal Jacobian  ADMM \label{alg1}}
}						
			\end{algorithm}
		\end{minipage}
		\vspace{-.2in}
\end{figure}
Note that we do not discuss the optimality of the proposed algorithm, in this paper, because the optimality of PJADMM was proven in \cite{deng2017parallel} when the objective function is a sum of strictly convex separable functions, which is the case in our objective function.
\\

\textbf{ Complexity analysis of the proposed algorithm}

We evaluate the computational complexity of the proposed algorithm based on the number of multiplication operations needed to detect the transmitted symbol vector. We adopt the number of real-valued multiplications for the analysis of the computational complexity as in \cite{qin2016near}, and \cite{yin2014conjugate}. The overall computational complexity of the PJADMM based algorithm consists of two parts: the first part, which is independent of the number of iterations, is needed to be performed once such as the multiplications operations of ${\sum_{k=1}^{2N_r}h_i^{(k)}}$, and the second part, which is iteration dependent, is need to be repeated in every iteration, such as the multiplication operation of ${\sum_{k=1}^{2N_r}h_i^{(k)}{y_i^{(k)}}^{(t-1)}}$. There is no computations required to perform initialization as the initial values of $x_i$ and $\textbf{y}$ can be all zeros. 

Based on ~\eqref{eq:x1Update}-\eqref{lambdaUpdate}, the iteration independent computations requires $4N_r$ real-valued multiplications, and the iteration dependent requires $T(14Nr+2N_t)$ real-valued multiplications. As it is shown in the next section, the number of iteration, $T$, depends on the number of uplink user antennas, $N_t$. Therefore, based on the parallel implementation of PJ ADMM, the number of time units required to process the detection of one element $\hat{x}_i$ is:
 
\begin{equation}
\# \text{of time units}=4N_r + T (14 Nr + 2 N_t)
\label{multiplication_operation}
\end{equation}
%$t_p=4N_r + T (14 Nr + 2 N_t)$
and since all elements of the received vector ($x_i, \forall i$ ) are processed in parallel, the total processing time for detecting one received vector is the same as that needed for detecting one element.

\section{Numerical Results}

To evaluate the performance of the proposed algorithm we implemented several simulation experiments for the uplink massive MIMO system in a block flat fading channel. We assume perfect knowledge of the channel state information at the receiver and uncoded QPSK modulations for demonstration, however the proposed algorithm can be extended for higher QAM modulations with coded cases. The aim of the simulation in this section is to show that the bit error rate (BER) performance of the proposed algorithm can achieve the performance of the exact MMSE technique (which is the benchmark \cite{dai2015low}) at various number of users with faster processing time. 

Regarding a comparison with other state of the art work, it is important to note that, all previous works such as \cite{wu2014large}, \cite{jiang2017low}, \cite{qin2016near}, \cite{jiang2018low}, assume the MMSE matrix ($\textbf{W}=\textbf{H}^H \textbf{H} + \sigma^2 \textbf{I}_{N_t}$) is a diagonal dominant matrix. This can be an acceptable assumption when the number of BS antennas is an order of magnitude more than the number of users. %, however, which cannot be always true, especially when the ratio between the number of BS antennas and the number of single antenna users is close to 1. 
Therefore, the proposed algorithm in this paper is a versatile in the sense that it can provide the same performance of the exact MMSE performance at any number of users, in addition, to providing fast processing time. 

\begin{figure}[t!]
%%%%\begin{minipage}[b]{1.0\linewidth}
\centering
\includegraphics[width=8cm, height=5cm]{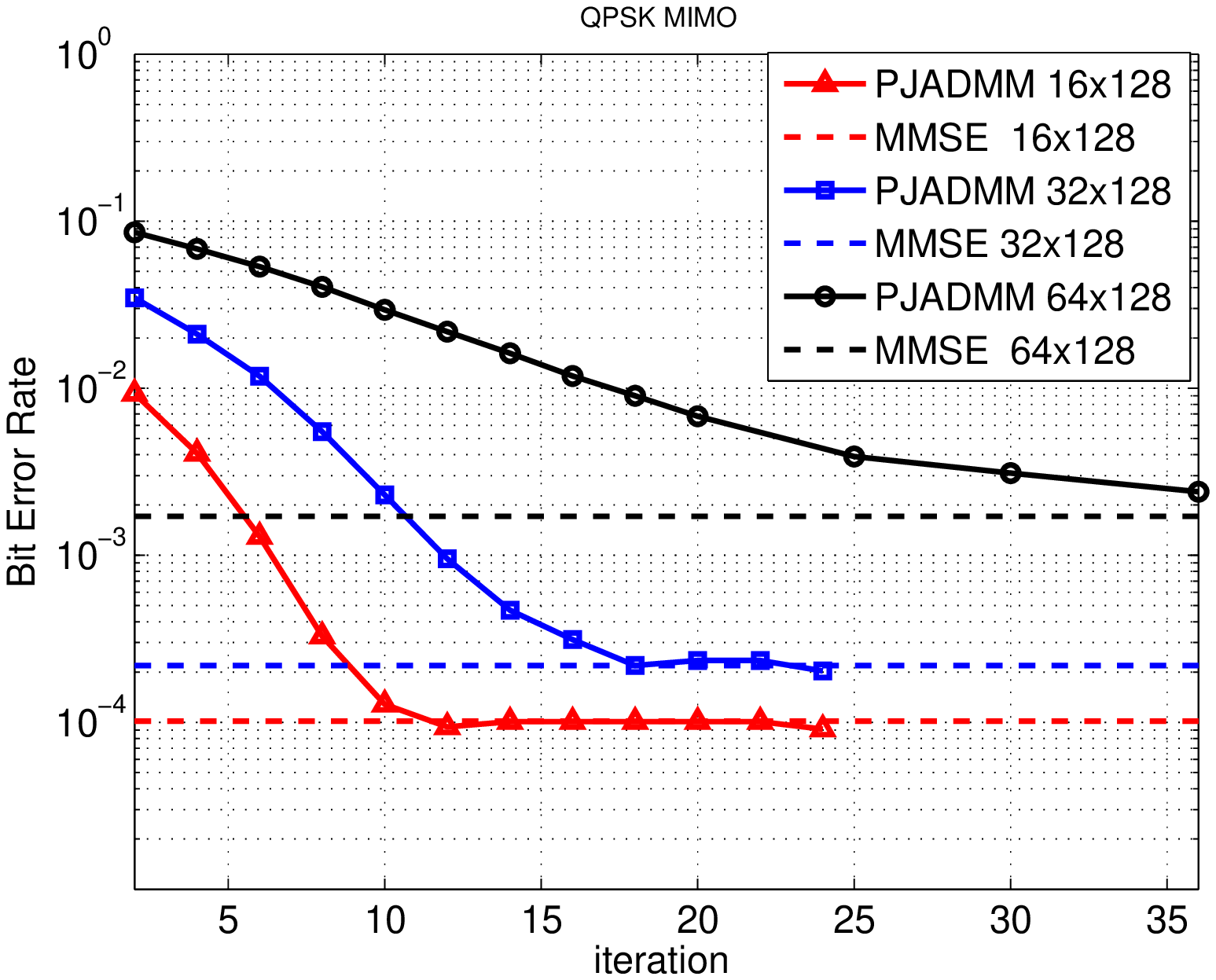}
\caption{\small  BER performance versus $\#$ of Iterations of the PJADMM based Algorithm for different massive MIMO configurations at $\textbf{SNR}=12$ dB }
\label{fig: 2}
\end{figure}

In the first simulation experiment, we studied the BER performance of our algorithm compared to the MMSE at various number of iterations and at fixed $\textbf{SNR}$ value. The number of iterations $T$ changes with a step size 2. The BER of various massive MIMO configurations are studied at $\textbf{SNR}=12$ dB. Fig. \ref{fig: 2} shows BER performance versus maximum number of iterations for the following MIMO configurations $16\times 128, 32 \times 128, 64\times 128 $. The MMSE performance is also shown in the same figure to examine the number of iterations required by the algorithm to reach MMSE performance. It can be seen from the figure that the larger the ratio between $N_r$ and $N_t$ the smaller the number of iterations required. The $16 \times 128$ configuration requires 12, while the $64 \times 128$ configuration requires 40 iterations to reach MMSE performance.

\begin{figure}[t!]
%%%%\begin{minipage}[b]{1.0\linewidth}
\centering
\includegraphics[width=8cm, height=5cm]{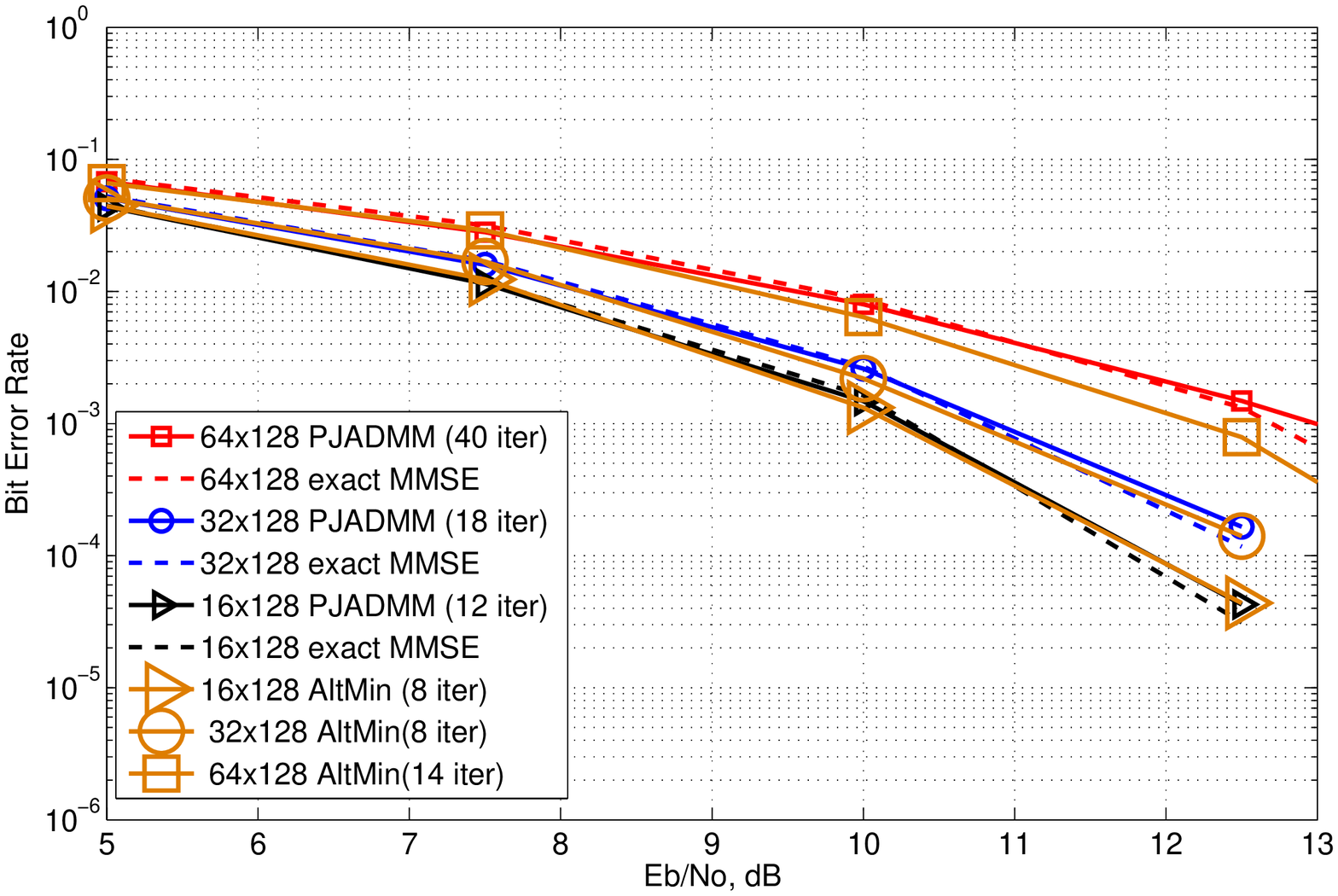}
\caption{\small BER performance of PJADMM based Algorithm at various massive MIMO configurations }
\label{fig: 1}
\end{figure}

%\textit{BER performance comparison}
Next, we use the number of iterations obtained in the pervious simulation to study the BER performance of our proposed algorithm compared to the exact MMSE performance at various $\textbf{SNR}$ values. Similar to the configuration above, the number of BS antennas is kept at 128, while the number of single antenna users can vary such that $N_t \leq N_r$, as depicted in Fig. \ref{fig: 1}. It can be clearly observed that the performance of our proposed algorithm matches that of the AltMin technique \cite{elgabli2018low} and the exact MMSE. Note, although the AltMin algorithm requires less iterations in each MIMO configuration, the proposed algorithm has the advantage of parallel detection implementations, i.e. it can provide lower latency. 
%the number of iteration required by the proposed algorithm is higher than the AltMIn algorithm, but the proposed al  It is also clear that as the number of users increases for a fixed number of BS antennas, the proposed algorithm requires more iterations. 

 %The proposed algorithm is examined to see its performance when the  number of  antennas increases. This important aspect in large MIMO detectors referred to as adherence to a large system behavior. It means that the performance of a MIMO detector increases as the number of antennas increases \cite{elg015quad}. Fig. \ref{fig: 2} shows that the BER performance of the ...... improves as $N_t \times N_r$ increases (e.g. , $ 8\times 8$, $ 16\times 16$, $ 32\times 32$, and $ 64\times 64$). For instance at BER= $10^{-5}$, the performance of $ 64\times 64$ is just about 1 dB away from SISO AWGN.
 
The next simulation experiment is used to emphasize the processing time of the proposed algorithm with its parallel implementation compared to the MMSE bench mark and also compared to a previous iterative techniques such as \cite{elgabli2018low}. The comparison is made in terms of the number of time units required to process a one received vector. We define the number of time units as the number of computations required, which is basically a function of the number of real-valued multiplication operations and also the number of iterations.The comparison is also based on the fact that all techniques have the same BER performance. \textcolor{black}{ Note, the aim of this paper is not to compare the performance of the proposed algorithm with other techniques such as those in  [11]-[13] because we prove that it achieves the exact MMSE performance as depicted in Fig. \ref{fig: 1}. However, we aim to provide a parallel implementation of the detection process using PJADMM that was not done in [11]-[13]}. Table \ref{compComplexity}, shows clearly how the proposed algorithm requires much less processing time compared to both the exact MMSE technique and the AltMin technique that has a successive symbol detection implementation. At a smaller number of users (i.e. the ratio between $N_r$ and $N_t$ is large) such as 16, the PJADMM based algorithm can provide a processing time that is two times faster than the exact MMSE and 10 times faster than AltMin. As $N_t $ increases with fixed $N_r$, the advantage of the proposed algorithm is of orders of magnitude faster. For example, at $N_t = 64$, it performs 18 times faster than AltMin and 28 times faster than exact MMSE.\\
\begin{table}[htp]
%	  \vspace{-.1in}
  \centering
  \caption{\small The number of time units ($\times$ $10^4$) required to process one received vector at $\textbf{SNR}=12$ dB}
  \begin{tabularx}{0.48\textwidth}{|X|X|X|X|X|} \hline
     & $N_t$=16 & $N_t$=32 & $N_t$=64 & $N_t$=128 \\ \hline
  \textbf{MMSE}  & 5.7 & 31.1 & 219.5 & 1697 \\ \hline
   \textbf{AltMin}\cite{elgabli2018low} & 20 & 40.9 & 140.9 & 281.8 \\ \hline
   \textbf{PJADMM}  & 2.24 & 3.39 & 7.73 & 10.3 \\ \hline
  \end{tabularx}
  \label{compComplexity}
  \vspace{-.1in}
\end{table}

% % % % % % % % % % % % % % % % % % % % % % % % % % % % % % % % % % % % % % % % %
%

%
%{\textit{{Turbo Coded BER Performance}} }: The turbo coded BER performance of the $Alt min$ algorithm compared to MMSE is shown in Fig. \ref{fig: 7.0} using coded QPSK modulation. In this simulation, all the above massive MIMO configurations are examined with rate-1/2 turbo encoder and decoder of 10 iterations. $ \pm1 $ output valued vector from all detectors is fed as an input to the BCJR-based turbo decoder. In Fig. \ref{fig: 7.0},  $Alt min$ algorithm performs similar to the MMSE detector for $16\times 128$, and slightly better than MMSE for $32\times 128$. As the number of uplink antennas increases, the coded $Alt min$ clearly outperforms coded MMSE, for example the improvement of $Alt min$ over MMSE for $128\times 128$ at $10^{-3}$  coded BER is about 1 dB compared to only 0.2 dB improvement in the case of $64\times 128$.

%\begin{figure}[h]
%%%%\begin{minipage}[b]{1.0\linewidth}
%\centering
%\includegraphics[width=8cm, height=6cm]{new_results/Coded_BER_MIMO_ntx128_alternate_algo.eps}
%\caption{\small 1/2 Turbo coded BER performance of QPSK for different massive MIMO configurations}
%\label{fig: 7.0}
%\end{figure}
%
%

%\input{simRes}

%

%\section{Future work}

%\vspace{-0.08in}

\section{Conclusion and Future work}

This paper presented a novel reformulation for the ML problem in such a way that it suits PJADMM optimization tool. The PJADMM technique is used to solve the new formulated problem in an iterative manner such that in every iteration all variables are updated in parallel and in a closed form expression, without a need for any matrix inversion. This provides much faster processing time compared to the exact MMSE and the other iterative-based techniques, especially when the number of single antenna users is close to the number of base station (BS) antennas. 
%\vspace{-0.150in}
%We are currently investigating Multi-Stage LASSO ADMM Signal Detection Algorithm For Large Scale MIMO where $N$ stages are considered $(N > 2)$. Our preliminary results show further performance improvement when the rounding thresholds are well chosen per stage and for some values of $\lambda$ which interests us to understand in more depth the theory behind choosing these parameters..Finally,  we are working on Multi-Stage Generalized LASSO ADMM Signal Detection Algorithm for Spatial Modulation. In spatial modulation, not only the coefficients are sparse, but also the transmitted symbols since only one or few antennas is/are active at any given time.
\bibliographystyle{IEEEtran}

\bibliography{refMIMO}

\end{document}